\documentclass[11pt]{article}
\usepackage{amsfonts}
 \usepackage{amsmath}

\usepackage{mathrsfs}
\usepackage{fancyhdr}
\usepackage{cite}
\newtheorem{proposition}{Proposition}[section]
\newtheorem{lemma}[proposition]{Lemma}

\begin{document}
\renewcommand{\theequation}{\thesection.\arabic{equation}}

\begin{center}
\LARGE\bf Finite genus solutions to the lattice Schwarzian Korteweg-de Vries equation  
\end{center}

\begin{center}
\rm Xiaoxue Xu$^{\rm a}$, Cewen Cao$^{\rm a}$, Guangyao Zhang$^{\rm
b}$
\end{center}

\begin{center}

${}^{\rm a}$ School of Mathematics and Statistics, Zhengzhou University, Zhengzhou, 450001, PR China\\
${}^{\rm b}$ School of Science, Huzhou University, Zhejiang, 313000, PR China
\end{center}

\begin{center}
E-mail:\ \ xiaoxuexu@zzu.edu.cn
\end{center}

\begin{center}\textbf{Abstract}
\end{center}
\begin{quote} {\quad \quad Based on integrable Hamiltonian systems related to the derivative Schwarzian Korteweg-de Vries (SKdV) equation, a novel discrete Lax pair for the lattice SKdV (lSKdV) equation is given by two copies of a Darboux transformation which can be used to derive an integrable symplectic correspondence. Resorting to the discrete version of Liouville-Arnold theorem, finite genus solutions to the lSKdV equation are calculated through Riemann surface method.}
\end{quote}



\section{Introduction}\setcounter{equation}{0}
Remarkable progress has been made in recent years in the study of
discrete soliton equations (see \cite{Hietarinta} and the references therein). Among the
related mathematical theories, the property of multi-dimensional
consistency plays an important role in the understanding of
discrete integrability. In the 2-dimensional case, it leads to the well-known Adler-Bobenko-Suris
(ABS) list \cite{Grammaticos,Adler}, which gives
a classification of integrable quadrilateral lattice equations. Quite a few works have appeared in
the study of the ABS equations, concerning their relations with the usual soliton equations, the
Lax pairs, explicit analytic solutions, B$\mathrm{\ddot{a}}$cklund transformations (BTs), symmetries and conservation laws etc.
\cite{Atkinson,Cao,Cao0,Hietarinta0,Nijhoff,Nijhoff0,Petrera,Rasin,Viallet,Xenitidis,Zhang}.

The purpose of this paper is to investigate the lSKdV equation, which was first given in \cite{Nijhoff1},
\begin{equation}{\label{eq1}}
\Xi:=\gamma_1^2(u-\hat{u})(\tilde{u}-\hat{\tilde{u}})-
\gamma_2^2(u-\tilde{u})(\hat{u}-\hat{\tilde{u}})=0,
\end{equation}
 where the usual notation is adopted:
$u=u(m,n)$, $\tilde{u}=u(m+1,n)$, $\hat{u}=u(m,n+1)$. Eq. (\ref{eq1}) is exactly the $\mathrm{Q1(0)}$ model ($\mathrm{Q1}$ with $\delta=0$) in the ABS hierarchy. The approach of Lax representation will be used to confirm the integrability of Eq. (\ref{eq1}) and to
calculate its basic explicit analytic solutions, the finite genus solutions \cite{Cao0,Cao}.

To produce a purely discrete Lax pair, it is vital to select two appropriate
discrete spectral problems. It turns out that a special role is played by the semi-discrete
integrable equations, which are also of independent interest, see \cite{Levi} and references
therein, where the well-known Toda, the Volterra and the Ablowitz-Ladik hierarchies are
investigated thoroughly. A semi-discrete Lax pair can be constructed with the help of a continuous spectral
problem and its Darboux transformation (DT), where the DT is regarded as a discrete spectral
problem \cite{Bruschi,Levi}, which usually leads to an integrable symplectic map by using the non-linearization technique \cite{Cao0,Cao,Cao1}. Refer to \cite{Kuznetsov}, integrable maps are called BTs whose geometrical explanation is given in terms of spectral curves and their Jacobians. And the symplectic correspondences (BTs) compatible with finite gap solutions of KdV have been discussed through DTs for the standard KdV spectral problem \cite{Hone}.

In our case we consider the continuous SKdV equation,
\begin{eqnarray}{\label{eq2}}
\displaystyle\frac{\phi_{y}}{\phi_{x}}+\displaystyle\frac{1}{4}S[\phi;x]=0,
\end{eqnarray}
where $S[\phi;x]$ denotes the Schwarzian derivative of $\phi$ \cite{Weiss0,Weiss,Lou}, i.e.
\begin{equation}{\label{eq3}}
\begin{split}
S[\phi;x]=&\Big(\displaystyle\frac{\phi_{xx}}{\phi_{x}}\Big)_{x}-\displaystyle\frac{1}{2}\Big(\displaystyle\frac{\phi_{xx}}{\phi_{x}}\Big)^{2},\\
=&\displaystyle\frac{\phi_{xxx}}{\phi_{x}}-\displaystyle\frac{3}{2}\Big(\displaystyle\frac{\phi_{xx}}{\phi_{x}}\Big)^{2}.
\end{split}
\end{equation}
Technically, it is more convenient to use the derivative version
\begin{eqnarray}{\label{eq4}}
w_{y}+\displaystyle\frac{1}{4}\Big(w_{xx}-\displaystyle\frac{3w^{2}_{x}}{2w}\Big)_{x}=0,
\end{eqnarray}
with $w=\phi_{x}$. The Eq. (\ref{eq4}) has a Lax pair given by
\begin{align}{\label{eq5}}
\partial_{x}\chi&=\begin{pmatrix}0&-w\lambda^{-1}\\w^{-1}\lambda^{-1}&0\end{pmatrix}\chi,\\
\partial_{y}\chi&=\begin{pmatrix}-\displaystyle \frac{w_{x}}{2w}\lambda^{-2}&-w\lambda^{-3}+\displaystyle\frac{1}{4}\Big(w_{xx}-\displaystyle\frac{3w^{2}_{x}}{2w}\Big)\lambda^{-1}\\
w^{-1}\lambda^{-3}+\displaystyle\frac{1}{4w^{2}}\Big(w_{xx}-\displaystyle\frac{w_{x}^{2}}{2w}\Big)\lambda^{-1}&\displaystyle \frac{w_{x}}{2w}\lambda^{-2}\end{pmatrix}\chi.{\label{eq6}}
\end{align}
We note that in \cite{Nijhoff2}, the Lax pair for one Schwarzian PDE, which is equivalent to the SKdV hierarchy via expansions on the independent variables and has a fully discrete counterpart (\ref{eq1}) by considering the independent variables as lattice parameters, has been found. However, we have not been able to blend it with the algebro-geometric technique of nonlinearization employed in the present paper. Fortunately, here each of the linear systems (\ref{eq5}) and (\ref{eq6}) can be nonlinearized to produce an integrable Hamiltonian system.
Thus we find a Liouville integrable system associated with a spectral problem (see Sec. 2) given by
\begin{align}{\label{eq7}}
&\partial_x\chi=U(\lambda,u)\chi,\quad
U(\lambda,u)=\left(\begin{array}{cc}u&\lambda\\0&-u\end{array}
\right),
\end{align}
and find that the following DT of Eq. (\ref{eq7}) is critical:
\begin{align}{\label{eq8}}
&\tilde{\chi}=(\lambda^2-\gamma^2)^{-1/2}D^{(\gamma)}(\lambda,b)\chi,\quad
D^{(\gamma)}(\lambda,b)=\left(\begin{array}{cc}\lambda&\gamma
b\\\gamma b^{-1}&\lambda\end{array}\right).
\end{align}
The compatibility condition
$D_x^{(\gamma)}=\tilde{U}D^{(\gamma)}-D^{(\gamma)}U$ gives rise to
\begin{equation}{\label{eq9}}
b_x/b=u+\tilde{u},\quad \gamma b^{-1}=u-\tilde{u}.
\end{equation}
This suggests a constraint $b=\gamma/(u-\tilde{u})$ and leads to a
Lax pair, different from the one in \cite{Nijhoff1}, for Eq. (\ref{eq1}).

\begin{lemma}
The lSKdV equation (\ref{eq1}) has a Lax pair
\begin{equation}\begin{split}{\label{eq10}}
&\tilde{\chi}=(\lambda^2-\gamma^2_1)^{-1/2}D^{(\gamma_1)}(\lambda,b^\prime)\chi,\quad
b^\prime=\gamma_1/(u-\tilde{u}),\\
&\hat{\chi}=(\lambda^2-\gamma^2_2)^{-1/2}D^{(\gamma_2)}(\lambda,b^{\prime\prime})\chi,\quad
b^{\prime\prime}=\gamma_2/(u-\hat{u}),
\end{split}\end{equation}
with
\begin{equation}{\label{eq11}}
\hat{D}^{(\gamma_1)}D^{(\gamma_2)}-\tilde{D}^{(\gamma_2)}D^{(\gamma_1)}=
\frac{1}{\Upsilon}\left(\begin{array}{cc}(u-\tilde{u})(u-\hat{u})&
\lambda(\hat{\tilde{u}}-\tilde{u}-\hat{u}+u)\\
0&-(\tilde{u}-\hat{\tilde{u}})(\hat{u}-\hat{\tilde{u}})\end{array}
\right)\Xi,
\end{equation}
where
$\Upsilon=(u-\tilde{u})(u-\hat{u})(\tilde{u}-\hat{\tilde{u}})(\hat{u}-\hat{\tilde{u}})$
and $\Xi$ is defined by Eq. (\ref{eq1}).
\end{lemma}

The paper is organised as follows. In Sec. 2, a finite-dimensional Hamiltonian system which is a nonlinear version of the spectral problem (\ref{eq7}) is presented. In Sec. 3, resorting to the Hamiltonian system, we construct an integrable symplectic map. In addition, with the help of the Burchnall-Chaundy theory, the discrete potential is expressed in terms of theta functions. In Sec. 4, based on the discrete version of the Liouville-Arnold theorem, the finite genus solutions of lSKdV equation (\ref{eq1}) are obtained through the commutativity of integrable maps \cite{Cao}.

\section{The integrable Hamiltonian system $(H_1)$}\setcounter{equation}{0}
Take the symplectic manifold $(\mathbb R^{2N},\mathrm{d}p \wedge \mathrm{d}q)$ as the
phase space. The symplectic coordinate is defined as $(p,q)=(p_{1},\ldots,p_{N},q_{1},\ldots,q_{N})$. Let $A=\textrm{diag}(\alpha_1,\cdots,\alpha_N)$ with
distinct, non-zero $\alpha_1^2,\cdots,\alpha_N^2$. Define a Lax
matrix
\begin{equation}{\label{eq12}}
L(\lambda;p,q)=\sigma_+
+\frac{1}{2}\sum_{j=1}^{N}\Big(\frac{\varepsilon_j}{\lambda-\alpha_j}
+\frac{\sigma_3\varepsilon_j\sigma_3}{\lambda+\alpha_j}\Big)
=\left(\begin{array}{cc}\lambda
Q_\lambda(p,q)&1-Q_\lambda(Ap,p)\\Q_\lambda(Aq,q)&-\lambda
Q_\lambda(p,q)\end{array}\right),
\end{equation}
where $\sigma_+, \sigma_3$ are the usual Pauli matrices, and
\begin{equation*}
\varepsilon_j=\left(\begin{array}{cc}p_jq_j&-p_j^2\\q_j^2&-p_jq_j\end{array}\right), \ \ Q_\lambda(\xi,\eta)=\sum_{j=1}^N\frac{\xi_j\eta_j}{\lambda^2-\alpha_j^2},
\end{equation*}
$\forall (\xi,\eta)=(\xi_{1},\ldots,\xi_{N},\eta_{1},\ldots,\eta_{N})\in \mathbb{R}^{2N}$.

\noindent The generating function $\mathcal{F}_\lambda=\det L(\lambda;p,q)$ is a rational
function of the argument $\zeta=\lambda^2$,
\begin{equation}{\label{eq13}}
\mathcal{F}_\lambda(p,q)=\big(Q_\lambda(Ap,p)-1\big)Q_\lambda(Aq,q)-\lambda^2Q_\lambda^2(p,q).
\end{equation}
The expansion $\mathcal{F}_\lambda=\sum_{l=1}^\infty F_l\zeta^{-l}$ gives
rise to a set of quantities on phase space as
\begin{equation}\begin{split}{\label{eq14}}
&F_1=-<Aq,q>-<p,q>^2,\\
&F_l=-<A^{2l-1}q,q>+\sum_{j+k=l;\,\,j,\,k\geq1}<A^{2j-1}p,p><A^{2k-1}q,q>\\
&\hspace{2cm}-\sum_{j+k=l+1;\,\,j,\,k\geq1}<A^{2j-2}p,q><A^{2k-2}p,q>\,,
\end{split}\end{equation}
$(l=1,2,\cdots)$, where $<\xi,\eta>=\Sigma_{j=1}^N\xi_j\eta_j$.
Consider the Hamiltonian system $(H_1)$, defined by the Hamiltonian function
\begin{equation}\begin{split}{\label{eq15}}
&H_1=\frac{F_1}{2}=-\frac{1}{2}<Aq,q>-\frac{1}{2}<p,q>^2,\\
&\partial_x{p_j\choose q_j}={{-\partial H_1/\partial
q_j}\choose{\partial H_1/\partial
p_j}}=\left(\begin{array}{cc}<p,q>&\alpha_j\\0&-<p,q>\end{array}\right){p_j\choose
q_j}\,,\quad 1\leq j\leq N.
\end{split}\end{equation}
They are exactly $N$ copies of Eq. (\ref{eq7}) with distinct
$\lambda=\alpha_j$ and the constraint
\begin{equation}{\label{eq16}}
u=f_U(p,q)=<p,q>.
\end{equation}
In this context $(H_1)$ is called a non-linearization of the linear spectral problem (\ref{eq7}).

According to the Liouville-Arnold theory \cite{Arnold}, we shall discuss the coefficients $F_1,\cdots,F_N$ given by (\ref{eq14}) are first integrals of the phase flow with Hamiltonian function $H_1$, i.e., $\{F_j,H_1\}=0, (j=1,\ldots,N)$, where $\{\cdot,\cdot\}$ denotes the Poisson bracket on the phase space. The involution and functional independence between $F_1,\cdots,F_N$ guarantee that the Hamiltonian system $(H_1)$ is completely integrable.

\noindent Consider the Hamiltonian system $(\mathcal{F}_\lambda)$,
\begin{equation}\begin{split}{\label{eq17}}
&\frac{\mathrm{d}}{\mathrm{d}t_\lambda}{{p_j}\choose{q_j}}={{-\partial
\mathcal{F}_\lambda/\partial q_j}\choose{\partial \mathcal{F}_\lambda/\partial p_j}}
=W(\lambda,\alpha_j){{p_j}\choose{q_j}},\\
&W(\lambda,\mu)=\frac{2}{\lambda^2-\mu^2}\left(\begin{array}{cc}\lambda L^{11}(\lambda)&\mu L^{12}(\lambda)\\
\mu L^{21}(\lambda)&-\lambda
L^{11}(\lambda)\end{array}\right)=\frac{L(\lambda)}{\lambda-\mu}+\frac{\sigma_3L(\lambda)\sigma_3}{\lambda+\mu},
\end{split}\end{equation}
where $L(\lambda)$ is the abbreviation of $L(\lambda;p,q)$ and $L^{ij}(\lambda),i,j=1,2$ are entries of the matrix $L(\lambda)$.
Hence we obtain
$\mathrm{d}\varepsilon_j/\mathrm{d}t_\lambda=[W(\lambda,\alpha_j),\varepsilon_j]$,
where $[\cdot,\cdot]$ stands for the matrix commutator. Based on
this formula, it is easy to derive the following basic equation,
\begin{equation}{\label{eq18}}
\frac{\mathrm{d}}{\mathrm{d}t_\lambda}L(\mu)=[W(\lambda,\mu),L(\mu)],\quad\quad\forall\lambda,\mu\in\mathbb
C.
\end{equation}
As a corollary, we have
\begin{align}{\label{eq19}}
&\{\mathcal{F}_\mu,\mathcal{F}_\lambda\}=0,\quad\forall\lambda,\mu\in\mathbb C;\\
&\{F_j,F_k\}=0,\quad j,k=1,2,\cdots.{\label{eq20}}
\end{align}
Actually, by equation (\ref{eq18}),
$(\mathrm{d}/\mathrm{d}t_\lambda)L^2(\mu)=[W(\lambda,\mu),L^2(\mu)]$. Since
$L^2(\mu)=-I\mathcal{F}_\mu$, where $I$ is the identity matrix, we have
$\mathrm{d}\mathcal{F}_\mu/\mathrm{d}t_\lambda=0$. According to the definition of Poisson
bracket \cite{Arnold}, this is exactly  Eq. (\ref{eq19}), whose power series
expansion gives rise to Eq. (\ref{eq20}).

The generating function $\mathcal{F}_\lambda$ has a factorization
\begin{equation}{\label{eq21}}
\mathcal{F}_\lambda=F_1\frac{Z(\zeta)}{\alpha(\zeta)}=F_1\frac{R(\zeta)}{\zeta\alpha^2(\zeta)},
\end{equation}
with $\alpha(\zeta)=\Pi_{j=1}^{N}(\zeta-\alpha_j^2)$, $
Z(\zeta)=\Pi_{k=1}^{N-1}(\zeta-\zeta_k)$, $
R(\zeta)=\zeta\alpha(\zeta)Z(\zeta)$, where $F_1$ is given by
Eq. (\ref{eq14}). The spectral curve is defined as
\begin{equation}{\label{eq22}}
\mathcal R:\,\xi^2-R(\zeta)=0,
\end{equation}
which is hyperelliptic with genus $g=N-1$ and has two points at infinity, $\infty_+,\,\infty_-$. At the branch point $\mathfrak
o=(\zeta=0,\xi=0)$, $\mathcal R$ has a local coordinate $\lambda=\zeta^{1/2}$. The generic point on $\mathcal R$ is given
as
$$
\mathfrak p(\zeta)=\big(\zeta,\xi=\sqrt{R(\zeta)}\big),\quad
(\tau\mathfrak p)(\zeta)=\big(\zeta,\xi=-\sqrt{R(\zeta)}\big),
$$
where $\tau:\mathcal R\rightarrow\mathcal R$ is the hyperelliptic involution. The variables $\{\nu_j^2\}$ defined as the roots of the equation
\begin{equation}{\label{eq23}}
L^{21}(\lambda)=\sum_{j=1}^{N}\displaystyle\frac{\alpha_{j}q_{j}^{2}}{\lambda^{2}-\alpha_{j}^{2}}=<Aq,q>\frac{\mathfrak n
(\zeta)}{\alpha(\zeta)}=0,\quad \mathfrak n
(\zeta)=\prod_{j=1}^g(\zeta-\nu_j^2),
\end{equation}
give an elliptic coordinate system\cite{Lame}.
By Eq. (\ref{eq18}) we have
\begin{equation}{\label{eq24}}
\frac{\mathrm{d}}{\mathrm{d}t_\lambda}L^{21}(\mu)=2\big(W^{21}(\lambda,\mu)L^{11}(\mu)-W^{11}(\lambda,\mu)L^{21}(\mu)\big).
\end{equation}
Putting $\mu=\nu_k$, with $L^{11}(\nu_k)=\sqrt{-F_1\cdot
R(\nu_k^2)}\Big/\big(\nu_k\alpha(\nu_k^2)\big)$ from Eq. (\ref{eq21}), we
get the evolution of the elliptic variables along the
$\mathcal{F}_\lambda$-flow,
\begin{align}{\label{eq25}}
&\frac{1}{2\sqrt{R(\nu_k^2)}}\cdot
\frac{\mathrm{d}(\nu_k^2)}{\mathrm{d}t_\lambda}=-\frac{2\sqrt{-F_1}}{\alpha(\zeta)}\cdot
\frac{\mathfrak n(\zeta)}{(\zeta-\nu_k^2)\mathfrak
n^{\prime}(\nu_k^2)},\quad 1\leq k\leq g,\\
&\sum_{k=1}^g\frac{(\nu_k^2)^{g-s}}{2\sqrt{R(\nu_k^2)}}\cdot
\frac{\mathrm{d}(\nu_k^2)}{\mathrm{d}t_\lambda}=-\frac{2\sqrt{-F_1}}{\alpha(\zeta)}\cdot\zeta^{g-s},\quad
1\leq s\leq g,{\label{eq26}}
\end{align}
where the interpolation formula of polynomials is used. With the
help of the quasi-Abel-Jacobi variables
\begin{equation}{\label{eq27}}
\phi_s^\prime=\sum_{k=1}^g\int_{\mathfrak p_0}^{\mathfrak
p(\nu_k^2)}\omega_s^{\prime},\quad
\omega_s^{\prime}=\frac{\zeta^{g-s}\mathrm{d}\zeta}{2\sqrt{R(\zeta)}},\quad
1\leq s\leq g,
\end{equation}
Eq. (\ref{eq26}) is rewritten in a simple form and gives rise to
\begin{proposition}
The $\mathcal{F}_\lambda$- and the $F_l$-flow are linearized by
$\phi_s^\prime$ as
\begin{align}{\label{eq28}}
&\frac{\mathrm{d}\phi_s^\prime}{\mathrm{d}t_\lambda}=\{\phi_s^\prime,\mathcal{F}_\lambda\}=-\frac{2\sqrt{-F_1}}{\alpha(\zeta)}\cdot\zeta^{g-s},\quad
1\leq s\leq g,\\
&\frac{\mathrm{d}\phi_s^\prime}{\mathrm{d}t_l}=\{\phi_s^\prime,F_l\}=-2\sqrt{-F_1}\cdot
A_{l-s-1},\quad
 l=1,2,\cdots,{\label{eq29}}
\end{align}
where $A_0=1$; $A_{-j}=0,\,(j=1,2,\cdots)$; while $A_j$,
$(j=1,2,\cdots)$, are defined by
$$
\frac{\zeta^N}{\alpha(\zeta)}=\frac{1}{\Pi_{k=1}^N(1-\alpha_k^2\zeta^{-1})}=\sum_{j=0}^{\infty}A_j\zeta^{-j}.
$$
In particular, $\{\phi_s^\prime,F_1\}=0,\,1\leq s\leq g$.
\end{proposition}
\begin{proposition}
The Hamiltonian system $(H_1)$ is integrable, possessing $N$
integrals $F_1,\cdots,F_N$, involutive with each other and
functionally independent in the dense, open subset $\mathcal
O=\{(p,q)\in\mathbb R^{2N}:\,F_1\neq0\}$.
\end{proposition}
\noindent\emph{Proof}. $F_l$ is an integral since
$\{H_1,F_l\}=(1/2)\{F_1,F_l\}=0$ by Eq. (\ref{eq20}). It needs only to prove that
$\mathrm{d}F_1,\cdots,\mathrm{d}F_N$ are linearly independent in
$T_{(p,q)}^\ast\mathbb R^{2N}$ at $(p,q)\in\mathcal O$. Suppose
$\Sigma_{j=1}^Nc_j\mathrm{d}F_j=0$. Then
$$
c_2\{\phi_s^\prime,F_2\}+\cdots+c_N\{\phi_s^\prime,F_N\}=0,\quad 1\leq
s\leq N-1.
$$
By Eq. (\ref{eq29}), the coefficient matrix is non-degenerate,
$$
\left(\begin{array}{ccc}\{\phi_1^\prime,F_2\}&\cdots&\{\phi_1^\prime,F_N\}\\\vdots&\ddots&\vdots\\
\{\phi_g^\prime,F_2\}&\cdots&\{\phi_g^\prime,F_N\}\end{array}\right)=-2\sqrt{-F_1}\cdot
\left(\begin{array}{ccccc}1&A_1&A_2&\cdots&A_{g-1}\\&1&A_1&\cdots&A_{g-2}\\&&\ddots&\cdots&\vdots\\
&&&1&A_1\\&&&&1\end{array}\right)\,\,.
$$
Thus $c_2=\cdots=c_N=0$ and $c_1\mathrm{d}F_1=0$. We have $c_1=0$ since
$\mathrm{d}F_1\neq 0$ at $\mathcal O$. Otherwise,
$$
-\frac{1}{2}\mathrm{d}F_1=\sum_{j=1}^{N}\big(<p,q>q_j\mathrm{d}p_j+(\alpha_jq_j+<p,q>p_j)\mathrm{d}q_j\big)=0.
$$
Hence $\alpha_jq_j+<p,q>p_j=0$, $\forall j$; and $F_1=0$. This is
a contradiction. \hfill $\Box$

\section{The integrable symplectic map $\mathcal S_\gamma$}\setcounter{equation}{0}
As a non-linearization of Eq. (\ref{eq8}), define a map $\mathcal
S_\gamma:\,\mathbb R^{2N}\rightarrow\mathbb
R^{2N},\,(p,q)\mapsto(\tilde{p},\tilde{q})$ by
\begin{equation}{\label{eq30}}
{{\tilde{p}_j}\choose{\tilde{q}_j}}=(\alpha_j^2-\gamma^2)^{-1/2}D^{(\gamma)}(\alpha_j,b){{p_j}\choose{q_j}},
\quad 1\leq j\leq N,
\end{equation}
where a constraint $b=f_\gamma(p,q)$ is to be chosen so that
$\mathcal S_\gamma$ is integrable and symplectic.

\begin{lemma}
Let
$P^{(\gamma)}(b;p,q)=b^2L^{21}(\gamma)+2bL^{11}(\gamma)-L^{12}(\gamma)$.
Then
\begin{align}{\label{eq31}}
&L(\lambda;\tilde{p},\tilde{q})D^{(\gamma)}(\lambda,b)-D^{(\gamma)}(\lambda,b)L(\lambda;p,q)
=-\gamma b^{-1}P^{(\gamma)}(b;p,q)\sigma_3,\\
&\sum_{j=1}^N(\mathrm{d}\tilde{p}_j\wedge \mathrm{d}\tilde{q}_j-\mathrm{d}p_j\wedge
\mathrm{d}q_j)=\frac{1}{2}\gamma b^{-2}\mathrm{d}P^{(\gamma)}(b;p,q)\wedge \mathrm{d}b.{\label{eq32}}
\end{align}
\end{lemma}
\noindent \emph{Proof}. By Eq. (\ref{eq30}), we get $
\tilde{\varepsilon}_jD^{(\gamma)}(\alpha_j)-D^{(\gamma)}(\alpha_j)\varepsilon_j=0.
$ Besides, we have $\sigma_3^{2}=I$ and
\begin{align*}
&\sigma_3D^{(\gamma)}(\lambda)\sigma_3=-D^{(\gamma)}(-\lambda),\notag\\
&D^{(\gamma)}(\pm\lambda)-D^{(\gamma)}(\alpha_j)=\pm(\lambda\mp\alpha_j)I.\notag
\end{align*}
Based on these preparations, we calculate the left-hand side of
Eq. (\ref{eq31}),
\begin{equation*}\begin{split}
[\sigma_{+},D^{(\gamma)}(\lambda)]
&+\frac{1}{2}\sum_{j=1}^N\Big(\frac{\tilde{\varepsilon}_jD^{(\gamma)}(\lambda)
-D^{(\gamma)}(\lambda)\varepsilon_j}{\lambda-\alpha_j}+
\frac{\sigma_3\tilde{\varepsilon}_j\sigma_3D^{(\gamma)}(\lambda)
-D^{(\gamma)}(\lambda)\sigma_3\varepsilon_j\sigma_3}{\lambda+\alpha_j}\Big)\\
&=\gamma
b^{-1}\sigma_3+\frac{1}{2}\sum_{j=1}^N\Big((\tilde{\varepsilon}_j-\varepsilon_j)
+\sigma_3(\tilde{\varepsilon}_j-\varepsilon_j)\sigma_3\Big)\\
&=(\gamma b^{-1}+<\tilde{p},\tilde{q}>-<p,q>)\sigma_3.\notag
\end{split}\end{equation*}
By using Eq. (\ref{eq30}), we obtain
\begin{equation} {\label{eq33}}
\gamma b^{-1}+<\tilde{p},\tilde{q}>-<p,q>=-\gamma
b^{-1}P^{(\gamma)}(b;p,q).
\end{equation}
This proves Eq. (\ref{eq31}). Eq. (\ref{eq32}) is obtained through direct
calculations. \hfill $\Box$

Consider the quadratic equation $P^{(\gamma)}(b)=0$, whose roots give the constraint on $b$,
\begin{equation}{\label{eq34}}
b=f_\gamma(p,q)=\frac{1}{Q_\gamma(Aq,q)}\Big(-\gamma
Q_\gamma(p,q)\pm\sqrt{-\mathcal{F}_\gamma(p,q)}\Big).
\end{equation}
Actually $\gamma b$ can be written as a meromorphic function on $\mathcal{R}$,
\begin{equation*}
\mathfrak{b}(\mathfrak{p})=\frac{1}{Q_\gamma(Aq,q)}\Big(-\gamma^{2}
Q_\gamma(p,q)+\sqrt{-F_1}\displaystyle\frac{\xi}{\alpha(\gamma)}\Big).
\end{equation*}
Though doubled-valued as a function of $\beta\in \mathbb{C}$, it is single-valued as a function of $\mathfrak{p}(\beta^{2})\in \mathcal{R}$. Hence we obtain
\begin{proposition}
The map $\mathcal S_\gamma:\mathbb R^{2N}\rightarrow\mathbb
R^{2N},\,(p,q)\mapsto(\tilde{p},\tilde{q})$, defined as
\begin{equation}{\label{eq35}}
{{\tilde{p}_j}\choose{\tilde{q}_j}}=(\alpha_j^2-\gamma^2)^{-1/2}{{\alpha_jp_j+\gamma
bq_j }\choose{\gamma
b^{-1}p_j+\alpha_jq_j}}\Big|_{b=f_\gamma(p,q)},\quad 1\leq j\leq
N,
\end{equation}
is symplectic and integrable, possessing the Liouville set of
integrals
\begin{equation}{\label{eq36}}
F_l(\tilde{p},\tilde{q})=F_l(p,q),\quad 1\leq j\leq N.
\end{equation}
\end{proposition}
\noindent\emph{Proof}. Since $P^{(\gamma)}(b)=0$, by Eq. (\ref{eq31}) and (\ref{eq32}) we have
\begin{align}{\label{eq37}}
&L(\lambda;\tilde{p},\tilde{q})D^{(\gamma)}\big(\lambda,f_\gamma(p,q)\big)
-D^{(\gamma)}\big(\lambda,f_\gamma(p,q)\big)L(\lambda;p,q)=0,\\
&\sum_{j=1}^N\mathrm{d}\tilde{p}_j\wedge
\mathrm{d}\tilde{q}_j=\sum_{j=1}^N\mathrm{d}p_j\wedge \mathrm{d}q_j.{\label{eq38}}
\end{align}
Taking the determinant of Eq. (\ref{eq37}), we obtain
$\mathcal{F}_\lambda(\tilde{p},\tilde{q})=\mathcal{F}_\lambda(p,q)$, hence Eq. (\ref{eq36}). \hfill $\Box$

By Eq. (\ref{eq36}), the discrete flow $\big(p(m),q(m)\big)=\mathcal
S_\gamma^m(p_0,q_0)$ has constants of motion $\{F_l\}$. Define
finite genus potentials as
\begin{align}{\label{eq39}}
&b(m)=b_m=f_\gamma\big(p(m),q(m)\big),\\
&u(m)=u_m=f_U\big(p(m),q(m)\big)=<p(m),q(m)>.{\label{eq40}}
\end{align}
By Eq. (\ref{eq33}), they have the relation
\begin{equation}{\label{eq41}}
b_m=\gamma/(u_m-u_{m+1}),
\end{equation}
which meets the requirement of Eq. (\ref{eq10}). Along the $m$-flow,
Eq. (\ref{eq37}) is rewritten as
\begin{equation}{\label{eq42}}
L_{m+1}(\lambda)D_m^{(\gamma)}(\lambda)=D_m^{(\gamma)}(\lambda)L_m(\lambda),
\end{equation}
where $L_m(\lambda)=L\big(\lambda;p(m),q(m)\big)$,
$D_m^{(\gamma)}(\lambda)=D^{(\gamma)}(\lambda,b_m)$. Now we
calculate $u_m$ with the help of the following spectral problem
and its fundamental solution matrix $M_\gamma(m,\lambda)$,
\begin{align}{\label{eq43}}
&h_\gamma(m+1,\lambda)=D_m^{(\gamma)}(\lambda)h_\gamma(m,\lambda);\\
&M_\gamma(m+1,\lambda)=D_m^{(\gamma)}(\lambda)M_\gamma(m,\lambda),\quad
M_\gamma(0,\lambda)=I.{\label{eq44}}
\end{align}
By induction we have
\begin{equation}\begin{split}{\label{eq45}}
&M_\gamma(m,\lambda)=D_{m-1}^{(\gamma)}(\lambda)D_{m-2}^{(\gamma)}(\lambda)\cdots D_0^{(\gamma)}(\lambda),\\
&\det M_\gamma(m,\lambda)=(\lambda^2-\gamma^2)^m,\\
&L_m(\lambda)M_\gamma(m,\lambda)=M_\gamma(m,\lambda)L_0(\lambda).
\end{split}\end{equation}

\begin{lemma}
The following functions are polynomials of the argument
$\zeta=\lambda^2$,
\begin{equation}\begin{split}{\label{eq46}}
&M_\gamma^{11}(2k,\lambda),\,\lambda^{-1}M_\gamma^{12}(2k,\lambda),\,
\lambda^{-1}M_\gamma^{21}(2k,\lambda),\,M_\gamma^{22}(2k,\lambda),\\
&\lambda^{-1}M_\gamma^{11}(2k+1,\lambda),\,M_\gamma^{12}(2k+1,\lambda),\,
M_\gamma^{21}(2k+1,\lambda),\,\lambda^{-1}M_\gamma^{22}(2k+1,\lambda).
\end{split}\end{equation}
Besides, as $\lambda\rightarrow\infty$,
\begin{equation}{\label{eq47}}
M_\gamma(m,\lambda)=\left(\begin{array}{cc}\lambda^m[1+O(\lambda^{-2})]&O(\lambda^{m-1})\\
O(\lambda^{m-1})&\lambda^m[1+O(\lambda^{-2})]\end{array}\right).
\end{equation}
\end{lemma}

By Eq. (\ref{eq42}), the solution space $\mathcal E_\lambda$ of Eq. (\ref{eq43})
is invariant under the action of the linear operator
$L_m(\lambda)$, which has two eigenvalues
$\rho_\lambda^\pm=\pm\rho_\lambda$,
\begin{equation}{\label{eq48}}
\rho_\lambda=\sqrt{-\mathcal{F}_\lambda}=\sqrt{-F_1}\cdot\frac{\sqrt{R(\zeta)}}{\lambda\alpha(\zeta)}.
\end{equation}
They define a meromorphic function $\mathfrak r(\mathfrak
p)=\sqrt{-F_1}\,\,\xi(\mathfrak p)/\alpha(\zeta(\mathfrak p))$ on
$\mathcal R$ with $\mathfrak r\big(\mathfrak
p(\lambda^2)\big)=\lambda\rho_\lambda^+$, $\mathfrak
r\big((\tau\mathfrak p)(\lambda^2)\big)=\lambda\rho_\lambda^-$.
The corresponding eigenvectors satisfy
\begin{align}{\label{eq49}}
&h_\pm(m,\lambda)={{h_\pm^{(1)}(m,\lambda)}\choose{h_\pm^{(2)}(m,\lambda)}}=M_\gamma(m,\lambda)
{{c_\lambda^\pm}\choose{1}},\\
&\big(L_m(\lambda)-\rho_\lambda^\pm\big)h_\pm(m,\lambda)=0.{\label{eq50}}
\end{align}
Putting $m=0$, we solve
\begin{align}{\label{eq51}}
&c_\lambda^\pm=\frac{L_0^{11}(\lambda)\pm\rho_\lambda}{L_0^{21}(\lambda)}=-\frac{L_0^{12}(\lambda)}
{L_0^{11}(\lambda)\mp\rho_\lambda}\,\,,\\
&c_\lambda^+c_\lambda^-=-\frac{L_0^{12}(\lambda)}{L_0^{21}(\lambda)},{\label{eq52}}
\end{align}
defining a meromorphic function $\mathfrak c(\mathfrak p)$ with
$\mathfrak c\big(\mathfrak p(\lambda^2)\big)=\lambda c_\lambda^+$,
$\mathfrak c\big((\tau\mathfrak p)(\lambda^2)\big)=\lambda
c_\lambda^-$. As $\lambda\rightarrow\infty$,
\begin{equation}{\label{eq53}}
c_\lambda^\pm=\frac{<p,q>\pm\sqrt{-F_1}}{<Aq,q>}\Big|_{<p_0,q_0>}\lambda[1+O(\lambda^{-2})].
\end{equation}

\begin{lemma}(Formula of Dubrovin-Novikov type.)
\begin{align}{\label{eq54}}
&\left(\begin{array}{cc}h_+^{(1)}h_-^{(1)}&h_+^{(1)}h_-^{(2)}\\h_+^{(2)}h_-^{(1)}&h_+^{(2)}h_-^{(2)}
\end{array}\right)\Bigg|_{(m,\lambda)}=\frac{(\lambda^2-\gamma^2)^m}{L_0^{21}(\lambda)}
\left(\begin{array}{cc}-L_m^{12}(\lambda)&L_m^{11}(\lambda)+\rho_\lambda\\
L_m^{11}(\lambda)-\rho_\lambda&L_m^{21}(\lambda)\end{array}\right),\\
&h_+^{(2)}(m,\lambda)h_-^{(2)}(m,\lambda)=\frac{<Aq,q>_m}{<Aq,q>_0}(\zeta-\gamma^2)^m\prod_{j=1}^g
\frac{\zeta-\nu_j^2(m)}{\zeta-\nu_j^2(0)}.{\label{eq55}}
\end{align}
\end{lemma}

\noindent\emph{Proof}. Using Eq. (\ref{eq45}), we calculate the
left-hand side of Eq. (\ref{eq54}),
\begin{align*}
LHS&=M_\gamma(m,\lambda)\left(\begin{array}{cc}c_\lambda^+c_\lambda^-&c_\lambda^+\\
c_\lambda^-&1\end{array}\right)M_\gamma^T(m,\lambda)\notag\\
&=\frac{1}{L_0^{21}(\lambda)}M_\gamma(m,\lambda)[L_0(\lambda)+\rho_\lambda
I]\left(\begin{array}{cc}0&1\\
-1&0\end{array}\right)M_\gamma^T(m,\lambda)\notag\\
&=\frac{1}{L_0^{21}(\lambda)}[L_m(\lambda)+\rho_\lambda
I]M_\gamma(m,\lambda)\left(\begin{array}{cc}0&1\\
-1&0\end{array}\right)M_\gamma^T(m,\lambda)\notag\\
&=\frac{1}{L_0^{21}(\lambda)}[L_m(\lambda)+\rho_\lambda
I]\left(\begin{array}{cc}0&1\\
-1&0\end{array}\right)\det M_\gamma(m,\lambda)=RHS.
\end{align*}
With the help of Eq. (\ref{eq23}) and (\ref{eq54}), Eq. (\ref{eq55}) is verified by some calculations. \hfill $\Box$
\begin{lemma}
As $\lambda\rightarrow\infty$,
\begin{align}{\label{eq56}}
&h_\pm^{(1)}(m,\lambda)=\frac{<p,q>_0\pm\sqrt{-F_1}}{<Aq,q>_0}\,\lambda^{m+1}[1+O(\lambda^{-2})],\\
&h_\pm^{(2)}(m,\lambda)=\frac{<p,q>_m\mp\sqrt{-F_1}}{<p,q>_0\mp\sqrt{-F_1}}\,\lambda^m[1+O(\lambda^{-2})].{\label{eq57}}
\end{align}
\end{lemma}

\noindent\emph{Proof}. Since
$h_\pm^{(1)}(m,\lambda)=M_\gamma^{11}(m,\lambda)c_\lambda^\pm+M_\gamma^{12}(m,\lambda)$,
we have Eq. (\ref{eq56}) in virtue of Eq. (\ref{eq47}) and (\ref{eq53}). By Eq. (\ref{eq54})
we get
\begin{align*}
&h_+^{(1)}h_-^{(2)}\big|_{(m,\lambda)}=(\lambda^2-\gamma^2)^m\frac{L_m^{11}(\lambda)+\rho_\lambda}{L_0^{21}(\lambda)}
=\frac{<p,q>_m+\sqrt{-F_1}}{<Aq,q>_0}\,\lambda^{2m+1}[1+O(\lambda^{-2})],\notag\\
&h_+^{(2)}h_-^{(1)}\big|_{(m,\lambda)}=(\lambda^2-\gamma^2)^m\frac{L_m^{11}(\lambda)-\rho_\lambda}{L_0^{21}(\lambda)}
=\frac{<p,q>_m-\sqrt{-F_1}}{<Aq,q>_0}\,\lambda^{2m+1}[1+O(\lambda^{-2})].\notag
\end{align*}
Thus we obtain Eq. (\ref{eq57}) by solving $h_\pm^{(2)}$ and using
Eq. (\ref{eq56}). \quad \quad \quad \quad \quad \quad \quad \quad \quad \quad \quad $\Box$

From Eq. (\ref{eq49}) we have
\begin{align*}
&h_\pm^{(2)}(2k,\lambda)=(\lambda
c_\lambda^\pm)\lambda^{-1}M_\gamma^{21}(2k,\lambda)+M_\gamma^{22}(2k,\lambda),\notag\\
&\lambda h_\pm^{(2)}(2k+1,\lambda)=(\lambda
c_\lambda^\pm)M_\gamma^{21}(2k+1,\lambda)+\lambda
M_\gamma^{22}(2k+1,\lambda).\notag
\end{align*}
By Lemma 3.3 and the discussion on $\lambda c_\lambda^\pm$, two
meromorphic functions (the Baker functions) $H^{(2)}(2k,\mathfrak
p)$ and $H^{(2)}(2k+1,\mathfrak p)$ are defined on $\mathcal R$,
respectively, with
\begin{equation}\begin{split}{\label{eq58}}
&H^{(2)}\big(2k,\mathfrak
p(\lambda^2)\big)=h_+^{(2)}(2k,\lambda),\quad
H^{(2)}\big(2k,(\tau\mathfrak
p)(\lambda^2)\big)=h_-^{(2)}(2k,\lambda),\\
&H^{(2)}\big(2k+1,\mathfrak p(\lambda^2)\big)=\lambda
h_+^{(2)}(2k+1,\lambda),\quad H^{(2)}\big(2k+1,(\tau\mathfrak
p)(\lambda^2)\big)=\lambda h_-^{(2)}(2k+1,\lambda).
\end{split}\end{equation}

\begin{proposition}
$H^{(2)}(2k,\mathfrak p)$ and $H^{(2)}(2k+1,\mathfrak p)$ have the
divisors respectively,
\begin{equation}\begin{split}{\label{eq59}}
&\Sigma_{j=1}^g[\mathfrak p\big(\nu_j^2(2k)\big)-\mathfrak
p\big(\nu_j^2(0)\big)]+2k\mathfrak
p(\gamma^2)-k(\infty_++\infty_-),\\
&\Sigma_{j=1}^g[\mathfrak p\big(\nu_j^2(2k+1)\big)-\mathfrak
p\big(\nu_j^2(0)\big)]+(2k+1)\mathfrak p(\gamma^2)+\mathfrak o
-(k+1)(\infty_++\infty_-).
\end{split}\end{equation}
\end{proposition}

\noindent\emph{Proof}. From Eq. (\ref{eq55}) and (\ref{eq58}) we obtain
\begin{equation}\begin{split}{\label{eq60}}
&H^{(2)}(2k,\mathfrak p)H^{(2)}(2k,\tau\mathfrak
p)=\frac{<Aq,q>_{2k}}{<Aq,q>_0}(\zeta-\gamma^2)^{2k}\prod_{j=1}^g\frac{\zeta-\nu_j^2(2k)}{\zeta-\nu_j^2(0)}\,\,,\\
&H^{(2)}(2k+1,\mathfrak p)H^{(2)}(2k+1,\tau\mathfrak
p)=\frac{<Aq,q>_{2k+1}}{<Aq,q>_0}\zeta(\zeta-\gamma^2)^{2k+1}\prod_{j=1}^g\frac{\zeta-\nu_j^2(2k+1)}{\zeta-\nu_j^2(0)}\,\,,
\end{split}\end{equation}
where $\mathfrak p=\mathfrak p(\zeta)$. As $\mathfrak
p\rightarrow\infty_\pm$, by Eq. (\ref{eq57}) and (\ref{eq58}) we have
\begin{equation}\begin{split}{\label{eq61}}
&H^{(2)}(2k,\mathfrak
p)=\frac{<p,q>_{2k}\mp\sqrt{-F_1}}{<p,q>_0\mp\sqrt{-F_1}}\,\,\zeta^k[1+O(\zeta^{-1})],\\
&H^{(2)}(2k+1,\mathfrak
p)=\frac{<p,q>_{2k+1}\mp\sqrt{-F_1}}{<p,q>_0\mp\sqrt{-F_1}}\,\,\zeta^{k+1}[1+O(\zeta^{-1})].
\end{split}\end{equation}
By these formulas it is easy to calculate the divisors. \quad \quad \quad \quad \quad \quad \quad \quad \quad \quad \quad \quad \quad \quad \quad \quad $\Box$

By using the technique developed by Toda \cite{Toda}, based on the
meromorphic differentials $\mathrm{d}\ln H^{(2)}(2k,\mathfrak p )$ and
$\mathrm{d}\ln H^{(2)}(2k+1,\mathfrak p )$, immediately we get
\begin{equation}\begin{split}{\label{eq62}}
&\sum_{j=1}^g\int_{\mathfrak p(\nu_j^2(0))}^{\mathfrak
p(\nu_j^2(2k))}\vec{\omega}+k\Big(\int_{\infty_+}^{\mathfrak
p(\gamma^2)}\vec{\omega}+\int_{\infty_-}^{\mathfrak
p(\gamma^2)}\vec{\omega}\Big)\equiv0,\quad(\textrm{mod}\,\,\mathcal
T),\\
&\sum_{j=1}^g\int_{\mathfrak p(\nu_j^2(0))}^{\mathfrak
p(\nu_j^2(2k+1))}\vec{\omega}+(k+1)\Big(\int_{\infty_+}^{\mathfrak
p(\gamma^2)}\vec{\omega}+\int_{\infty_-}^{\mathfrak
p(\gamma^2)}\vec{\omega}\Big)+\int_{\mathfrak
p(\gamma^2)}^{\mathfrak
o}\vec{\omega}\equiv0,\quad(\textrm{mod}\,\,\mathcal T),
\end{split}\end{equation}
where $\vec{\omega}=(\omega_1,\cdots,\omega_g)^T$ are the
normalized basis of holomorphic differentials on $\mathcal R$,
while $\mathcal T$ is the basic lattice spanned by the periodic
vectors of $\mathcal R$ \cite{Griffiths,Farkas}. With the help of the Abel map
$\mathcal A: \textrm{Div}(\mathcal R)\rightarrow J(\mathcal R)$,
$\mathcal A(\mathfrak p)=\int_{\mathfrak p_0}^{\mathfrak
p}\vec{\omega}$, the Abel-Jacobi variable is defined as
\begin{equation}{\label{eq63}}
\phi(m)=\mathcal A\Big(\sum_{j=1}^g\mathfrak
p\big(\nu_j^2(m)\big)\Big).
\end{equation}
This endows Eq. (\ref{eq62}) with a clear geometric explanation.

\begin{proposition}
In the Jacobi variety $J(\mathcal R)=\mathbb C^g/{\mathcal T}$,
the discrete flow $\mathcal S_\gamma^m$ is linearized by the
Abel-Jacobi variable
\begin{equation}{\label{eq64}}
\phi(m)\equiv\phi(0)+m\Omega_\gamma+\delta_m\Omega_{0\gamma},\quad
(\textrm{mod}\,\,\mathcal T),
\end{equation}
where $\delta_{2k}=0\,,\delta_{2k+1}=1$, and
\begin{equation}{\label{eq65}}
\Omega_\gamma=\frac{1}{2}\Big(\int^{\infty_+}_{\mathfrak
p(\gamma^2)}\vec{\omega}+\int^{\infty_-}_{\mathfrak
p(\gamma^2)}\vec{\omega}\Big),\quad\Omega_{0\gamma}=\Omega_\gamma+\int_{\mathfrak
o}^{\mathfrak p(\gamma^2)}\vec{\omega}.
\end{equation}
\end{proposition}

The meromorphic function $H^{(2)}(2k,\mathfrak p)$ is expressed by
its divisor up to a constant factor
\begin{equation}\begin{split}{\label{eq66}}
H^{(2)}(2k,\mathfrak p)=&const\cdot\frac{\theta[-\mathcal
A(\mathfrak p)+\phi(2k)+K]}{\theta[-\mathcal A(\mathfrak
p)+\phi(0)+K]}\\&\cdot\exp\Big\{k\int_{\mathfrak p_0}^{\mathfrak
p}\omega[\mathfrak p(\gamma^2),\infty_+]+\omega[\mathfrak
p(\gamma^2),\infty_-]\Big\},
\end{split}\end{equation}
where $K$ is the Riemann constant vector and $\omega[\mathfrak
p,\mathfrak q]$ is an Abel differential of the third kind, possessing
only two simple poles at $\mathfrak p,\mathfrak q$ with residues
$+1,-1$, respectively. Resorting to Eq. (\ref{eq61}), by the asymptotic
behaviors of Eq. (\ref{eq66}) near $\infty_\pm$ we obtain
\begin{equation}\begin{split}{\label{eq67}}
&\frac{u_{2k}-\sqrt{-F_1}}{u_0-\sqrt{-F_1}}=const\cdot\frac{\theta[-\mathcal
A(\infty_+)+\phi(2k)+K]}{\theta[-\mathcal
A(\infty_+)+\phi(0)+K]}(r_\gamma^+r_{\gamma,-}^+)^k,\\
&\frac{u_{2k}+\sqrt{-F_1}}{u_0+\sqrt{-F_1}}=const\cdot\frac{\theta[-\mathcal
A(\infty_-)+\phi(2k)+K]}{\theta[-\mathcal
A(\infty_-)+\phi(0)+K]}(r_\gamma^-r_{\gamma,+}^-)^k,
\end{split}\end{equation}
where
\begin{equation}{\label{eq68}}
r_\gamma^\pm=\lim_{\mathfrak
p\rightarrow\infty_\pm}\frac{1}{\zeta(\mathfrak
p)}\exp\int_{\mathfrak p_0}^{\mathfrak p}\omega[\mathfrak
p(\gamma^2),\infty_\pm],\quad
r_{\gamma,\,\mp}^\pm=\exp\int_{\mathfrak
p_0}^{\infty_\pm}\omega[\mathfrak p(\gamma^2),\infty_\mp].
\end{equation}
We introduce a new variable $v_m$ by
\begin{equation}{\label{eq69}}
v_m=\frac{u_m-\sqrt{-F_1}}{u_m+\sqrt{-F_1}}\,\,,\quad
u_m=\sqrt{-F_1}\frac{1+v_m}{1-v_m}\,\,.
\end{equation}
Cancelling the constant factor in Eq. (\ref{eq67}), we arrive at
\begin{equation}{\label{eq70}}
v_{2k}=v_0\cdot\frac{\theta[2k\Omega_\gamma+\Omega+K(0)]\cdot\theta[K(0)]}
{\theta[2k\Omega_\gamma+K(0)]\cdot\theta[\Omega+K(0)]}\cdot
e^{2kR_\gamma},
\end{equation}
where $\Omega=\int_{\infty_+}^{\infty_-}\vec{\omega}$ and
\begin{equation}\begin{split}{\label{eq71}}
&-\mathcal A(\infty_-)=\int_{\infty_-}^{\mathfrak
p_0}\vec{\omega}=\eta_-,\quad-\mathcal A(\infty_+)=\Omega+\eta_-,\\
&K(m)=\phi(m)+K+\eta_-,\quad
R_\gamma=\frac{1}{2}\ln[(r_\gamma^+r_{\gamma,-}^+)/(r_\gamma^-r_{\gamma,+}^-)].
\end{split}\end{equation}
Similarly, considering the analytic expression for
$H^{(2)}(2k+1,\mathfrak p)$ leads to
\begin{equation}{\label{eq72}}
v_{2k+1}=v_0\cdot\frac{\theta[(2k+1)\Omega_\gamma+\Omega_{0\gamma}+\Omega+K(0)]\cdot\theta[K(0)]}
{\theta[(2k+1)\Omega_\gamma+\Omega_{0\gamma}+K(0)]\cdot\theta[\Omega+K(0)]}\cdot
e^{(2k+1)R_\gamma+R_{0\gamma}},
\end{equation}
where
\begin{equation}{\label{eq73}}
R_{0\gamma}=R_\gamma+\ln r_{0\gamma},\quad
r_{0\gamma}=\exp\int_{\infty_-}^{\infty_+}\omega[\mathfrak
o,\mathfrak p(\gamma^2) ].
\end{equation}

\begin{proposition}
The finite genus potential $v_m$, defined by Eq. (\ref{eq40}) and (\ref{eq69}),
has an explicit evolution formula along the discrete flow
$\mathcal S_\gamma^m$ ,
\begin{equation}{\label{eq74}}
v_m=v_0\cdot\frac{\theta[m\Omega_\gamma+\delta_m\Omega_{0\gamma}+K(0)+\Omega]\cdot\theta[K(0)]}
{\theta[m\Omega_\gamma+\delta_m\Omega_{0\gamma}+K(0)]\cdot\theta[K(0)+\Omega]}\cdot
e^{mR_\gamma+\delta_mR_{0\gamma}},
\end{equation}
where the vectors $K(m),\,\Omega_\gamma,\,\Omega_{0\gamma}$ and $
\Omega$ are given by Eq. (\ref{eq65}) and (\ref{eq71}), while the constants
$R_\gamma,R_{0\gamma}$ are defined by Eq. (\ref{eq71}) and (\ref{eq73});
moreover, $\delta_{2k}=0$, $\delta_{2k+1}=1$, for all $k$.
\end{proposition}

\section{Solutions of lSKdV equation (\ref{eq1})}\setcounter{equation}{0}
Let $\gamma_1,\gamma_2$ be the two constants given in Eq. (\ref{eq1}).
By proposition 3.2, setting $\gamma=\gamma_1,\gamma_2$ in the above we have two symplectic maps
$\mathcal S_{\gamma_1}$ and $\mathcal S_{\gamma_2}$, sharing the same set of integrals $\{F_l\}$. Resorting to the
discrete version of Liouville-Arnold theorem \cite{Quispel,Suris,Veselov}, they commute. Thus we have
well-defined functions with two discrete arguments $m$ and $n$,
\begin{equation}\begin{split}{\label{eq75}}
&\big(p(m,n),q(m,n)\big)=\mathcal S_{\gamma_1}^m\mathcal
S_{\gamma_2}^n(p_0,q_0),\\
&b_{mn}=f_\gamma\big(p(m,n),q(m,n)\big),\\
&u_{mn}=f_U\big(p(m,n),q(m,n)\big)=<p(m,n),q(m,n)>,\\
&v_{mn}=(u_{mn}-\sqrt{-F_1}\,\,)\big/(u_{mn}+\sqrt{-F_1}\,\,).
\end{split}\end{equation}

\begin{proposition}
Both the functions $u_{mn}$ and $v_{mn}$, defined by Eq. (\ref{eq75}),
solve Eq. (\ref{eq1}).
\end{proposition}

\noindent\emph{Proof}. By the commutativity of $\mathcal
S_{\gamma_1}^m$ and $\mathcal S_{\gamma_2}^n$, we have
\begin{equation}{\label{eq76}}
\big(p(m,n),q(m,n)\big)=\mathcal
S_{\gamma_1}^m\big(p(0,n),q(0,n)\big)=\mathcal
S_{\gamma_2}^n\big(p(m,0),q(m,0)\big).
\end{equation}
From Eq. (\ref{eq41}) we obtain
\begin{equation}{\label{eq77}}
b_{mn}=\gamma_1/(u-\tilde{u})=\gamma_2/(u-\hat{u}).
\end{equation}
By Eq. (\ref{eq35}), $\chi_j=\big(p_j(m,n),q_j(m,n)\big)^T$ solves
simultaneously
\begin{equation}\begin{split}{\label{eq78}}
&\tilde{\chi}_j=(\alpha_j^2-\gamma_1^2)^{-1/2}D^{(\gamma_1)}(\alpha_j,b_{mn})\chi_j,\quad b_{mn}=\gamma_1/(u-\tilde{u}),\\
&\hat{\chi}_j=(\alpha_j^2-\gamma_2^2)^{-1/2}D^{(\gamma_2)}(\alpha_j,b_{mn})\chi_j,\quad
b_{mn}=\gamma_2/(u-\hat{u}).
\end{split}\end{equation}
Thus $u_{mn}$ satisfies Eq. (\ref{eq1}) by Eq. (\ref{eq11}). In order to prove
that $v_{mn}$ is also a solution, it is sufficient to notice that
$(i)$ $F_1$ is a constant of motion which is independent of $m$ and $n$;
$(ii)$ Eq. (\ref{eq1}) is invariant under the M\"{o}bius transformation
$u\mapsto v$ given by Eq. (\ref{eq75}). \hfill$\Box$

Apply Eq. (\ref{eq74}) to the flow $\mathcal S_{\gamma_1}^m$ and
$\mathcal S_{\gamma_2}^n$ successively. By $v_{00}\rightarrow
v_{m0}\rightarrow v_{mn}$ we obtain

\begin{proposition}
The lSKdV equation (\ref{eq1}) has finite genus solutions
\begin{equation}\begin{split}{\label{eq79}}
v_{mn}=v_{00}&\cdot\frac{\theta[m\Omega_{\gamma_1}+n\Omega_{\gamma_2}+\delta_m\Omega_{0\gamma_1}
+\delta_n\Omega_{0\gamma_2}+K_{00}+\Omega] \cdot\theta[K_{00}]}
{\theta[m\Omega_{\gamma_1}+n\Omega_{\gamma_2}+\delta_m\Omega_{0\gamma_1}
+\delta_n\Omega_{0\gamma_2}+K_{00}] \cdot\theta[K_{00}+\Omega]}\\
&\cdot
\exp(mR_{\gamma_1}+nR_{\gamma_2}+\delta_mR_{0\gamma_1}+\delta_nR_{0\gamma_2}),
\end{split}\end{equation}
and $u_{mn}=\sqrt{-F_1}\,\,(1+v_{mn})/(1-v_{mn})$. Further, any
M\"{o}bius transformation
$w_{mn}=(a_{11}v_{mn}+a_{12})/(a_{21}v_{mn}+a_{22})$ solves
Eq. (\ref{eq1}), where $a_{jk}$ are constants.
\end{proposition}

\textbf{Acknowledgment}. This work is supported by National Natural Science Foundation of China (Grant
Nos. 11426206; 11501521), State Scholarship Found of China (CSC No. 201907045035), and Graduate Student Education Research Foundation
of Zhengzhou University (Grant No. YJSXWKC201913). We would like to thank Prof. Frank W Nijhoff and Prof. Da-jun Zhang for helpful discussions.

\end{document}